\documentclass[aps,prl,twocolumn,amsmath,amssymb,showpacs]{revtex4}
\usepackage{graphicx}
\usepackage{Jonasmacros}
\usepackage{bm}

\begin{document}
\title{Angular conductance resonances of quantum dots non-collinearly coupled to ferromagnetic leads}
\author{J. Fransson}
\affiliation{Department of Materials Science and Engineering, Royal Institute of Technology (KTH), SE-100 44\ \ Stockholm, Sweden}
\affiliation{Physics Department, Uppsala University, Box 530, SE-751 21\ \ Uppsala, Sweden}
\affiliation{NORDITA, Blegdamsvej 17, DK-2100 Copenhagen, Denmark}

\begin{abstract}
The zero bias conductance of quantum dots coupled to ferromagnetic leads is investigated. In the strong coupling regime, it is found that the conductance is a non-monotonic function of the angle between the magnetisation directions in the two contacts. This behaviour is an effect of the presence of the leads which induces an angle dependent spin split of the quantum dot states, and spin flip transitions between the quantum dot states whenever the magnetisation directions of the leads are non-collinear which enhances the current density at the chemical potential. In the weak coupling regime, the system reverts to normal spin valve character.
\end{abstract}
\maketitle

Due to the fundamental complexities and far-reaching technological possibilities, the interest in spin-dependent transport in mesoscopic systems remain as high as ever. Since the first measurements of giant magneto resistance (GMR) in Fe/Cr magnetic super-lattices \cite{baibich1988}, magneto-transport has been studied for normal or ferromagnetic metallic islands \cite{metall_island}, and spin-dependent transport from ferromagnets through quantum dots (QDs) \cite{magnQD,konig2003} and molecules \cite{pati2003}. Recently, Kondo physics of QDs weakly coupled to ferromagnetic leads have been extensively studied \cite{martinek12003,martinek22003,choi2004}, showing that the suppression of the Kondo resonance due to the spin-polarisation of the leads may be restored by application of an external magnetic field.

In this Letter, effects of strong Coulomb interactions on the linear conductance is studied, in a single level QD coupled to ferromagnetic leads in the absence of external magnetic fields. The conductance of the system is predicted to display a huge complexity, for non-collinear alignment of the magnetic contacts. Due to the strong on-site correlations, the localised states undergo a spin-split depending on the non-collinear magnetisation of the leads, which in combination with effects from spin flip transitions, result in a non-monotonic conductance in the strong coupling regime. The QD system is analysed in terms of non-equilibrium many-body operator Green functions (GFs) \cite{franssonPRL2002,fransson2004} which is motivated for three reasons, namely, 1) the on-site Coulomb repulsion is the largest energy scale of the system, 2) a renormalisation of the QD level similar to the the scaling relation found in \cite{martinek12003} is included in the QD GF \cite{franssonPRL2002}, which cannot be obtained within traditional standard methods, and 3) the theory is valid in the whole range from the weak to the strong coupling regime \cite{franssonPNFA2004}. In the present case the Kondo effect is suppressed by the ferromagnetism in the leads \cite{martinek12003,martinek22003,choi2004}, and can therefore be neglected.

Consider a single level QD with a bare quantum level which is spin-degenerate, in the atomic limit, at the energy $\dote{0}$, and that the on-site Coulomb repulsion is given by $U$. Hence, the energy of the QD is given by $\Hamil_{QD}=\sum_\sigma\dote{0}\ddagger{\sigma}\d{\sigma}+Un_\up n_\down$, where $\ddagger{\sigma}\ (\d{\sigma})$ creates (annihilates) an electron in the QD, whereas $n_\sigma=\ddagger{\sigma}\d{\sigma}$, and $\sigma=\up,\down$ is the spin projection in the global reference frame. Since the Coulomb repulsion is the largest energy scale, the QD is rewritten in terms of its eigenstates, e.g. $\ket{p},\ p=0,\sigma,2$, where $\ket{2}=\ket{\up\down}$, hence $\Hamil_{QD}=\sum_pE_p\h{p}{}$, where $\h{p}{}=\X{pp}{},\ (\X{pq}{}=\ket{p}\bra{q})$, and $E_0=0$, $E_\sigma=\dote{0}$, and $E_2=2\dote{0}+U$ are the energies of the empty ($\ket{0}$), singly ($\ket{\sigma}$) and doubly ($\ket{2}$) occupied states, respectively.

\begin{figure}
\begin{center}
\includegraphics[width=8.5cm]{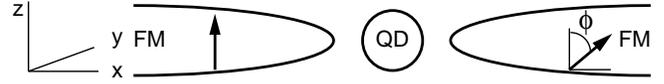}
\end{center}
\caption{Quantum dot in non-collinear coupling to the ferromagnetic leads. The co-ordinate system (left) defines the global reference frame. The magnetisation directions of the leads (arrows) enclose the angle $\phi$.}
\label{fig-system}
\end{figure}
The leads are modelled by $\Hamil_{L/R}=\sum_{k\sigma\in L/R}\leade{k}\cdagger{k}\c{k}$, where $\cdagger{k}\ (\c{k})$, creates (annihilates) an electron in the left/right ($L/R$) contact at the energy $\leade{k}$. The magnetisation in the left contact coincides with the $z$-axis of the global reference frame, see Fig. \ref{fig-system}, hence the tunnelling interaction between the QD and left contact is given by $\Hamil_{TL}=\sum_{k\sigma\in L,a}(v_{k\sigma}(\d{\sigma})^a\cdagger{k}\X{a}{}+H.c.)$, where $\sum_a(\d{\sigma})^a\X{a}{}=\bra{0}\d{\sigma}\ket{\sigma}\X{0\sigma}{}+\bra{\bar\sigma}\d{\sigma}\ket{2}\X{\bar\sigma2}{}$, and $\bar\sigma$ is the opposite spin of $\sigma$. In the following, it will assumed that the Coulomb repulsion $U$ is sufficiently large so that the doubly occupied state can be neglected, for briefness. Thus, $\sum_a(\d{\sigma})^a\X{a}{}=\X{0\sigma}{}$, since $\bra{0}\d{\sigma}\ket{\sigma}=1$, which yields $\Hamil_{TL}=\sum_{k\sigma\in L}(v_{k\sigma}\cdagger{k}\X{0\sigma}{}+H.c.)$. However, the doubly occupied state is easily included by a straight forward generalisation of the results presented below, and has been included into all numerical examples given. The magnetisation in the right contact is rotated by the angle $\phi$ in the global $xz$-plane, see Fig. \ref{fig-system}. Therefore, the tunnelling between the QD and the right contact is modelled by
\begin{eqnarray*}
\Hamil_{TR}=\sum_{k\in R} \biggl\{{\cal C}_k^T
	\left(\begin{array}{cc} v_{k\up} & 0 \\ 0 & v_{k\down}\end{array}\right)
	{\cal R}(\phi)
	\left(\begin{array}{c} \X{0\up}{} \\ \X{0\down}{} \end{array}\right)
	+H.c.\biggr\},
\end{eqnarray*}
where the vector ${\cal C}_k^T=(\csdagger{k+}\ \csdagger{k-})$ and the spin rotation matrix
\[ {\cal R}(\phi)=\left(\begin{array}{cc} \cos{\phi/2} & \sin{\phi/2} \\ 
	-\sin{\phi/2} & \cos{\phi/2} \end{array}\right).
\]
Here, the spin indices $s=\pm$ are used because of the rotated magnetisation in the right lead. Letting the magnetisation in the left lead coincide with the global $z$-axis is not a restriction since the magnetic properties of the QD depends on the angle between the magnetisation directions in the leads.

The total Hamiltonian for the system is given by $\Hamil=\sum_{\alpha=L,R}(\Hamil_\alpha+\Hamil_{T\alpha})+\Hamil_{QD}$ (stray fields from the leads are neglected). The spin-dependence of the leads is modelled by the hybridisation functions $\Gamma_\sigma^\alpha(\omega)=2\pi\sum_k|v_{k\sigma}^\alpha|^2\delta(\omega-\leade{k})=2\pi|v_\sigma^\alpha|^2\rho_\sigma^\alpha(\omega)$, where $v_{k\sigma}^\alpha\equiv v_\sigma^\alpha$, and $\rho_\sigma^\alpha$ is the spin-dependent density of states in the leads which is assumed to be constant. The spin-dependence of $\Gamma_\sigma^{\alpha}$ is parametrised in terms of $p_\alpha\equiv(\Gamma_\up^\alpha-\Gamma_\down^\alpha)/(\Gamma_\up^\alpha+\Gamma_\down^\alpha)$, letting $\Gamma_\sigma^\alpha=\Gamma_0(1\pm p_\alpha)$, where $2\Gamma_0=\Gamma_\up^\alpha+\Gamma_\down^\alpha$. By this procedure no essential physics is lost, as discussed in \cite{martinek22003}. In terms of the spin-dependent parameters $p_\alpha$, the coupling matrices to the left/right leads become
$\bfGamma^L=\Gamma_0\,\mbox{diag}\{1+p_L,1-p_L\}$ (diagonal matrix) and
\[ \bfGamma^R=\Gamma_0
	\left(\begin{array}{cc} 1+p_R\cos{\phi} & p_R\sin{\phi} \\
		p_r\sin{\phi} & 1-p_R\cos{\phi} \end{array}\right).
\]

A straight forward derivation shows that the zero bias conductance for this system can be written as \cite{meir1992} 
\begin{eqnarray}
G(\mu,\phi)=\frac{e^2}{h}\int\tr{}[ \bfGamma^L\bfG^r(\omega,\mu,\phi)\bfGamma^R\bfG^a(\omega,\mu,\phi)]
\nonumber\\
	\times\frac{\beta}{4}\cosh^{-2}\biggl(\beta\frac{\omega-\mu}{2}\biggr)d\omega
\label{eq-g}
\end{eqnarray}
for low temperatures $(\beta^{-1}=k_BT)$, where $\mu$ is the chemical potential of the system. In this expression, $\bfG^{r/a}(\omega,\mu,\phi)=\bfD^{r/a}(\omega,\mu,\phi)\bfP(\mu,\phi)$ is the retarded/advanced Green function (GF) of the localised states in the QD, where $\bfP(\mu,\phi)$ and $\bfD^{r/a}(\omega,\mu,\phi)$ denote the spectral weight and locator, respectively. The locator carries the local on-site properties of the GF, that is, the positions and widths of the poles. More details on the definition of the many-body operator GFs and the diagrammatic technique employed here can be found in \cite{franssonPRL2002,sandalov2003,fransson2004}. Following the steps in \cite{franssonPRL2002} and including the $\phi$ dependence of the coupling to the right, the QD GF is derived to (details will be published in \cite{fransson2005})
\begin{equation}
[\omega I-\bfDelta(\mu,\phi)-\bfSigma^{r/a}(\mu,\phi)]
	\bfG^{r/a}(\omega,\mu,\phi)=\bfP(\mu,\phi),
\label{eq-dyson}
\end{equation}
where $I$ is the identity matrix, whereas the transition energy matrix $\bfDelta(\mu,\phi)$ is renormalised by kinematic interactions between particles in the different localised states due to the presence of the de-localised electrons in the contacts. Here, the locator is identified by $\bfD^{r/a}(\omega,\mu,\phi)=[\omega I-\bfDelta(\mu,\phi)-\bfSigma^{r/a}(\omega,\mu,\phi)]^{-1}$. The transition energy matrix is given by
\begin{eqnarray}
\bfDelta(\mu,\phi)&=&\bfDelta^0+\sum_{\alpha=L,R}\int\frac{f_\alpha(\dote{})-f(\omega)}{\dote{}-\omega}
\nonumber\\&&
	\times\sigma_x\bfGamma^\alpha
	[-2\im\bfD^r(\omega,\mu,\phi)]\sigma_x
	\frac{d\omega}{2\pi} \frac{d\dote{}}{2\pi}.
\label{eq-delta}
\end{eqnarray}
where $\bfDelta^0=(\dote{0}-E_0)I$ is the bare energy matrix, $f_\alpha(\omega)=f(\omega-\mu_\alpha)$ is the Fermi function ($\mu_{L/R}$ is the chemical potential of the left/right lead; here $\mu_\alpha=\mu$), and $\sigma_x$ is the $x$-component of the Pauli spin vector. The self-energy in Eq. (\ref{eq-dyson}) is given by $\bfSigma^{r/a}(\mu,\phi)=\mp i\bfP(\mu,\phi)\sum_\alpha\Gamma^\alpha/2$, since the real part is negligible due to the large conduction electron band width in the leads. Apart from the renormalisation of the transition energies, Eq. (\ref{eq-dyson}) reduces to the result by Varma and Yafet \cite{varma1976} in the non-magnetic case for large $U$.

It should be emphasised that the QD GF has to be self-consistently calculated, for each point in the parameter space $(\mu,\phi,T,\Delta^0,\Gamma_0,p_L,p_R)$, along with Eq. (\ref{eq-delta}) and evaluation of the end-factor $\bfP$. The latter quantity is defined by $P_{\sigma\sigma'}=\Occu{\anticom{\X{0\sigma}{}}{\X{\sigma'0}{}}}=\delta_{\sigma\sigma'}N_0+N_{\sigma'\sigma}$, that is, as a sum of the population numbers of the states involved. These are calculated by $N_0=-\im\sum_\sigma\int G^>_{\sigma}(\omega)d\omega/(2\pi)$, $N_{\sigma}=\im\int G_{\sigma}(\omega)d\omega/(2\pi)$, and $N_{\sigma'\sigma}=-i\int G_{\sigma\sigma'}^<(\omega)d\omega/(2\pi)$ (note the reversed order of the spin indices), where $\bfG^{</>}=\bfG^r\bfV^{</>}\bfG^a$, $G_\sigma^{</>}$ and $G^{</>}_{\sigma\sigma'}$ are the diagonal and off-diagonal components of $\bfG^{</>}$, and $\bfV^<=i[f_L\bfGamma^L+f_R\bfGamma^R]$ and $\bfV^>=\bfV^<-i[\bfGamma^L+\bfGamma^R]$. The self-consistent calculations are subject to the condition $N_0+\sum_\sigma N_\sigma=1$, which satisfies the requirement that the integrated total QD density of states is unity, e.g. $\im\int\tr[\bfG^<-\bfG^>]d\omega/(2\pi)=1$. In this fashion effects from the spin-flip transitions are taken into account by self-consistently solving the full $2\times2$ matrix equation, e.g. Eq. (\ref{eq-g}), giving non-vanishing off-diagonal components whenever $0<\phi/\pi<1$.

The renormalisation of the transition energies, Eq. (\ref{eq-delta}), arise due to kinematic interactions between particles in the localised states induced by the presence of the de-localised electrons in the contacts. The magnetic properties of the contacts are transferred into the QD by this renormalisation, since the energy for the spin $\sigma$ state are influenced by the properties in the spin $\bar\sigma$ channel of the system. This is easiest seen for collinear leads, since then $\Sigma_{\sigma\bar\sigma},\ P_{\sigma\bar\sigma}=0$. Assuming zero width of the locator in Eq. (\ref{eq-delta}) and large band-width $2W\gg|\mu-\Delta_\sigma|$ of the conduction bands, result in \cite{franssonPRL2002}
\begin{equation}
\Delta_{\sigma}=\Delta^0+\sum_{\alpha=L/R}\frac{\Gamma_{\bar\sigma}^\alpha}{2\pi}
	\log{\biggl|\frac{\mu-\Delta_{\bar\sigma}}{W}\biggr|}.
\label{eq-simpdelta}
\end{equation}
\begin{figure}
\begin{center}
\includegraphics[width=8.5cm]{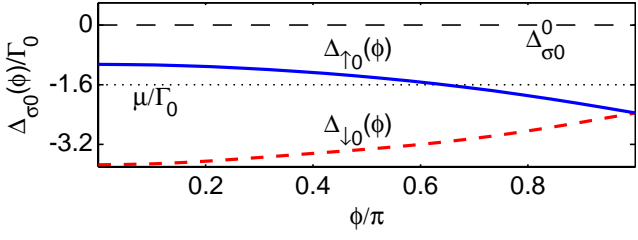}
\end{center}
\caption{(Colour online). Renormalised transition energies as function of $\phi$. Here $p_{L/R}=0.5$ and $\{\Delta^0,U,W,T\}/\Gamma_0=\{0,50,2000,0.17\}$.}
\label{fig-Delta}
\end{figure}

This expression is consistent with the results in \cite{varma1976,barabanov1974} for the non-magnetic case, as well as with the scaling equation reported in \cite{martinek12003} for the magnetic case. The logarithm is negative which leads to $\Delta_\sigma\leq\Delta^0$. By the replacement $\Delta_\sigma\rightarrow\Delta^0$ on the right hand side of Eq. (\ref{eq-simpdelta}), e.g. the first iteration in the self-consistent calculations, one finds the difference $\Delta_\up-\Delta_\down=-(\Gamma_0/\pi)(p_L+p_R\cos{\phi})\log{|(\mu-\Delta^0)/W|}$, in agreement with \cite{martinek12003}. Consequently, this difference is positive (negative) for $p_\alpha<0\ (p_\alpha>0)$, i.e. for parallel $(\phi=0)$ magnetic leads such that $\Gamma_\down^\alpha>\Gamma_\up^\alpha,\ (\Gamma_\down^\alpha<\Gamma_\up^\alpha)$, while for it vanishes for anti-parallel $(\phi=\pi)$ leads such that $p_L=p_R$. The renormalised transition energies in these two limits can be viewed in Fig. \ref{fig-Delta} ($\phi/\pi=0,\ 1$). The result in Eq. (\ref{eq-delta}) provides a continuous connection for the transition energies between the two collinear configurations of the leads, where the maximal and minimal spin split in the QD are given for the parallel and anti-parallel cases, respectively, as shown in Fig. \ref{fig-Delta}. The characteristics of the transition energies $\Delta_\sigma$ vary only slowly with $\mu$, and therefore the plots in Fig. \ref{fig-Delta} will be used independently of $\mu$.

\begin{figure}
\begin{center}
\includegraphics[width=8.5cm]{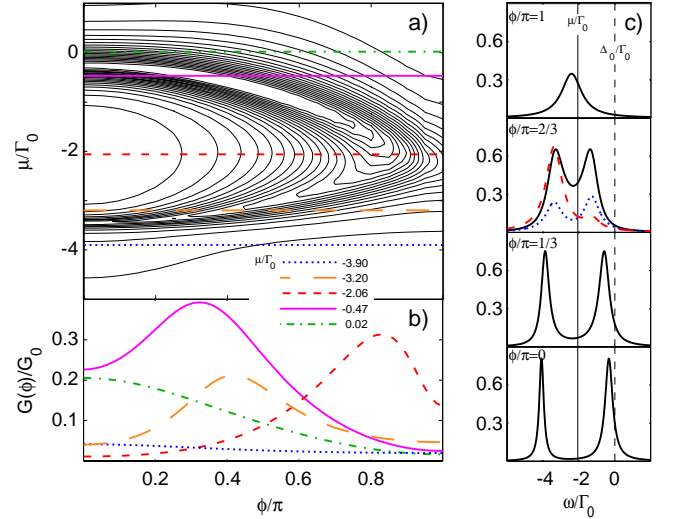}
\end{center}
\caption{(Colour online). a) Contour plot of the conductance, $0\leq G(\mu,\phi)/G_0\leq1/2$, where $G_0=2e^2/h$, as function of the chemical potential, $\mu$, and angle, $\phi$. b) The conductance, $G(\mu,\phi)/G_0$, for various values of the chemical potential indicated in panel a). c) Current density for different angles $\phi$ at $\mu/\Gamma_0=-2.06$ (straight solid). The second panel from above also shows the spin projected local DOS. Here $p_{L/R}=0.85$ and $\Delta^0/\Gamma_0=0$ (straight dashed).}
\label{fig-G}
\end{figure}
For non-collinear leads, e.g. $0<\phi/\pi<1$, the physics of the QD becomes more subtle, however, since the off-diagonal elements of $\bfGamma^R$ are finite, which leads to that the spin-flip transitions acquire non-vanishing weights, e.g. $N_{\sigma\bar\sigma}\neq0$, hence, the off-diagonal components of the QD GF $G_{\sigma\sigma'}(i\omega)\neq0$. Then, it is expected that both spin projections of the local density of states (DOS) $\rho_\sigma=\im[G^<_\sigma-G^>_\sigma]/(2\pi)=-\im{G_\sigma^r}/\pi$ become mixtures of one another in the sense that both $\rho_\up$ and $\rho_\down$ are peaked around both $\Delta_\up$ and $\Delta_\down$, since for instance
\[ G_\sigma^r(\omega)=
	\frac{(\omega-\Delta_{\bar\sigma}-\Sigma_{\bar\sigma}^r)P_{\sigma}
	+(\Delta_{\sigma\bar\sigma}+\Sigma_{\sigma\bar\sigma}^r)P_{\bar\sigma\sigma}}
	{(\omega-z_1^r)(\omega-z_2^r)}
\]
where the complex roots are given by $(n=1,2)$
\[ z_n^r=\lim_{\omega\rightarrow0}
	(\tr\bfD^{r,-1}+(-1)^n\sqrt{\tr^2\bfD^{r,-1}-4\det{\bfD^{r,-1}}})/2,
\]
and $\Delta_{\sigma\bar\sigma},\ \Sigma_{\sigma\bar\sigma}\neq0$. An example of the local DOS in this case is plotted in Fig. \ref{fig-G} c) for $\phi/\pi=2/3$, where it is readily seen that both $\rho_\up$ (dotted) and $\rho_\down$ (dashed) are doubly peaked, having finite densities in between.

In the remainder of this Letter, I discuss the effects of the $\phi$ dependent renormalised transition energies and QD GF on the zero bias conductance. The contour plot in Fig. \ref{fig-G} a) displays $G(\mu,\phi)$ as the systems is swept from weak coupling ($\mu-\max\{\Delta_\up,\Delta_\down\}\gg\Gamma_0$) to strong coupling and then back to weak coupling ($\mu-\min\{\Delta_\up,\Delta_\down\}\ll-\Gamma_0$), explicitly showing the complexity of the expected conductance in different regimes. Note that the character of the conductance peaks resembles the shape of the transition energies in Fig. \ref{fig-Delta}. In the weakly coupled regime, the QD effectively behaves as a spin-valve, where the conductance $G(\mu,\phi)$ is a monotonic function of $\phi$, see Fig. \ref{fig-G} b) (dotted, dash-dotted), in agreement with \cite{konig2003}. This is expected since the minimal distance $\min_\sigma|\mu-\Delta_\sigma|$ grows with $\phi$, which then leads to a reduction of the electron density in the QD around $\mu$.

In contrast, when $\mu$ lies in the vicinity of one or both transition energies (strong coupling), the qualitative character of the conductance is dramatically altered. In the parallel configuration ($\phi/\pi=0$) where the spin split of the QD transition energies is maximal, the two spin projected states, peaked at $\Delta_\up\neq\Delta_\down$, are separated roughly by $\Gamma_0(p_L+p_R)/\pi$, and with widths $P_\sigma\Gamma_0[2\pm(p_L+p_R)]/2$, since $\Sigma_{\sigma\bar\sigma},\ P_{\sigma\bar\sigma}=0$. For finite $\phi/\pi<1$, the two spin projected states become intermixed due to spin flip transitions, i.e. $\Sigma_{\sigma\bar\sigma},\ P_{\sigma\bar\sigma}\neq0$. This provides an enhancement of the electron density around $\mu$ which then is reflected in the current density $j(\omega,\mu,\phi)=\tr\bfGamma^L\bfG^r(\omega,\mu,\phi)\bfGamma^R\bfG^a(\omega,\mu,\phi)$, see Fig. \ref{fig-G} c), which displays $j(\omega,\mu,\phi)$ for various rotation angles. As seen in this plot, the amplitude of the current density at $\mu/\Gamma_0$ increases for growing rotation angles up to a threshold angle $\phi^*/\pi,\ 0<\phi^*/\pi<1$, and thereafter decreases.

\begin{figure}
\begin{center}
\includegraphics[width=8.5cm]{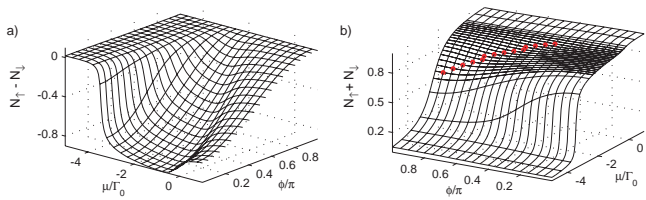}
\end{center}
\caption{(Colour online). a) Average spin-polarisation, $N_\up-N_\down$, and b) occupation, $N_\up+N_\down$, of the QD as functions of the chemical potential and angle. System parameters as in Fig. \ref{fig-G}. In b), the dots indicate the minimum occupation at the corresponding value of the chemical potential for rotation angles $0<\phi/\pi<1$.}
\label{fig-N}
\end{figure}
The effects of the $\phi$ dependent renormalised transition energies on the QD spin-polarisation, $N_\up-N_\down$, and occupation (valence), $N_\up+N_\down$, provides as additional understanding of the conductance. The plots in Fig. \ref{fig-N} display a) the calculated spin-polarisation and b) the occupation of the QD in the $(\mu,\phi)$-plane, and it is clearly seen that the QD is strongly spin-polarised for all values of the chemical potential between the QD state energies, i.e. $\min_\sigma\{\Delta_{\sigma0}\}<\mu<\max\{\Delta_{\sigma0}\}$, for $\phi/\pi=0$. The high QD occupation and large spin-polarisation in this regime reveals that there is only a small fraction of the local DOS available for transfer of electrons through the QD, since the spin-polarisation of the QD is opposite to that of the leads. As expected from the discussion of the spin-split of the QD state energies, the spin-polarisation decreases and eventually vanishes as $\phi/\pi\rightarrow1$. Due to the spin-split of the QD states, it is also clear that the spin-polarisation in the regime $\mu-\min_\sigma\{\Delta_{\sigma0}\}<0$ is smaller than in the regime $\mu-\max_\sigma\{\Delta_{\sigma0}\}>0$ since the QD is almost empty in the former regime whereas it is almost fully occupied in the latter, c.f. Fig. \ref{fig-N} b). As is signified by the marks in Fig. \ref{fig-N} b), the occupation of the QD clearly has a non-monotonic $\phi$ dependence in the strongly coupled regime $(\min_\sigma\{\Delta_{\sigma0}\}<\mu<\max_\sigma\{\Delta_{\sigma0}\})$. Although the characteristics of the occupation and the transport properties not are completely correlated throughout the whole $(\mu,\phi)$-plane, a closer analysis of the QD occupation, in the strongly coupled regime, reveals that its minimum occurs at the position of maximal conductance.

In conclusion, it is predicted that the linear conductance of a QD strongly coupled to ferromagnetic leads is a non-monotonic function of the angle between the magnetisation directions of the leads. The predicted character is an effect of an angle dependent spin split of the QD state, induced by the presence of the leads, and spin flip transitions between the QD states whenever the magnetisation directions of the leads are non-collinear. In accordance with previous considerations \cite{konig2003}, the system reverts to normal spin valve character in the weak coupling regime.

\acknowledgments
The author thanks O. Eriksson and O. Bengone for helpful discussions. Support from the G\"oran Gustafsson's foundation and Swedish Foundation for Strategic Research (SSF) is acknowledged.

\end{document}